%% file: CCNC2017.tex
\UseRawInputEncoding

\documentclass[conference]{IEEEtran}

\usepackage[dvips]{graphicx}

\ifCLASSINFOpdf
\else
\fi
\begin{document}
\title{Security Camera Movie and ERP Data Matching System to Prevent Theft}

\author{\IEEEauthorblockN{Yoji Yamato, Yoshifumi Fukumoto and Hiroki Kumazaki}
\IEEEauthorblockA{Software Innovation Center, NTT Corporation\\
Musashino-shi, Tokyo 180--8585, Japan\\
Email: \{yamato.yoji, fukumoto.yoshifumi, kumazaki.hiroki\}@lab.ntt.co.jp}}


%Telephone: (800) 555--1212, Fax: (888) 555--1212}


\maketitle

\begin{abstract}
In this paper, we propose a SaaS service which prevents shoplifting using image analysis and ERP. In Japan, total damage of shoplifting reaches 450 billion yen. Based on cloud and data analysis technology, we propose a shoplifting prevention service with image analysis of security camera and ERP data check for small shops. We evaluated movie analysis.
\end{abstract}



% \fi
\IEEEpeerreviewmaketitle

\input{section1E}
\input{section2E}
\input{section3E}
\input{section4E}
\input{section5E}

\input{reference}
\end{document}

%% file: section1E.tex
\section{Introduction}
Recently, cloud technology such as \cite{Yamato3}\cite{Yamato9}\cite{Yamato10}, service coordination technology such as \cite{Sunaga}\cite{Yamato6} and data analysis technology have been progressed. Big data analysis using Apache Spark or Hadoop with MapReduce achieves various analysis services.

On the other hand, total damage of shoplifting reaches 450 billion yen per year in Japan. To prevent shoplifting, stores adopt countermeasures such as increasing monitoring staffs in stores, checking security camera movie by human eyes or installing EAS (Electronic Article Surveillance) which alerts shoplifting at the gate of stores. However, these countermeasures need additional staff expense cost, initial cost of EAS or other systems. Thus, small shops cannot adopt them. 

Based on these backgrounds, this paper targets a low cost shoplifting prevention SaaS service for small shops using cloud technology and data analysis technology. In our proposal, machine learning framework Jubatus\cite{Jubatus} on a small computer deployed in a shop analyzes security cameras movie, detects anomaly behavior and notifies to a cloud. Then, a shoplifting prevention application on a cloud checks product stock using item DB of ERP and notifies smart phones of shop staffs by mails when a possibility of shoplifting is high.

%% file: section2E.tex
\section{Existing Technologies and Problems}
Saburo-kun Ace\cite{Saburo} is a shoplifting prevention system using security camera movie. Saburo-kun Ace detects a shoplifting from security camera movie when customers' actions match pre-defined 50 patterns of suspicious behaviors, and notifies it to staffs of shops. Shop staffs question or say something to the suspicious customer. This can reduce or prevent shopliftings. However, Saburo-kun Ace has some problems such as initial cost is high because shops need to deploy PC and movie analysis software, new shoplifting behaviors cannot be detected except for pre-defined suspicious behavior rules, actual operation may be hard because precision ratio of detection is not 100\% and shop staffs often need to question customers.

Existing technologies have two problems. The first is some technologies only can detect shoplifting based on pre-defined behavior rules. The second is accuracy of camera movie analysis is not sufficiently high so that actual operation may be difficult for staffs to question customers at unnecessary timing. Therefore, we target shoplifting prevention SaaS for small shops which detects shoplifting behavior including non-defined behavior at high accuracy and notifies shoplifting.

%% file: section3E.tex
\section{Proposal of Shoplifting Prevention Service Using Image Analysis and ERP Check}
To detect shoplifting of undefined behavior, we use online machine learning technology and detect suspicious actions comparing normal operation data. To enhance accuracy of shoplifting detection, we check not only camera movie analysis results but also product item DB managed on a cloud.

Figure 1 shows system image of proposed service. In our system, shop site which has security cameras and cloud site which manages product item data are connected by a network. Using Fig.1, we explain processing steps of proposed service.

 \begin{figure}[tb]
 \begin{center}
  \includegraphics[width=84mm]{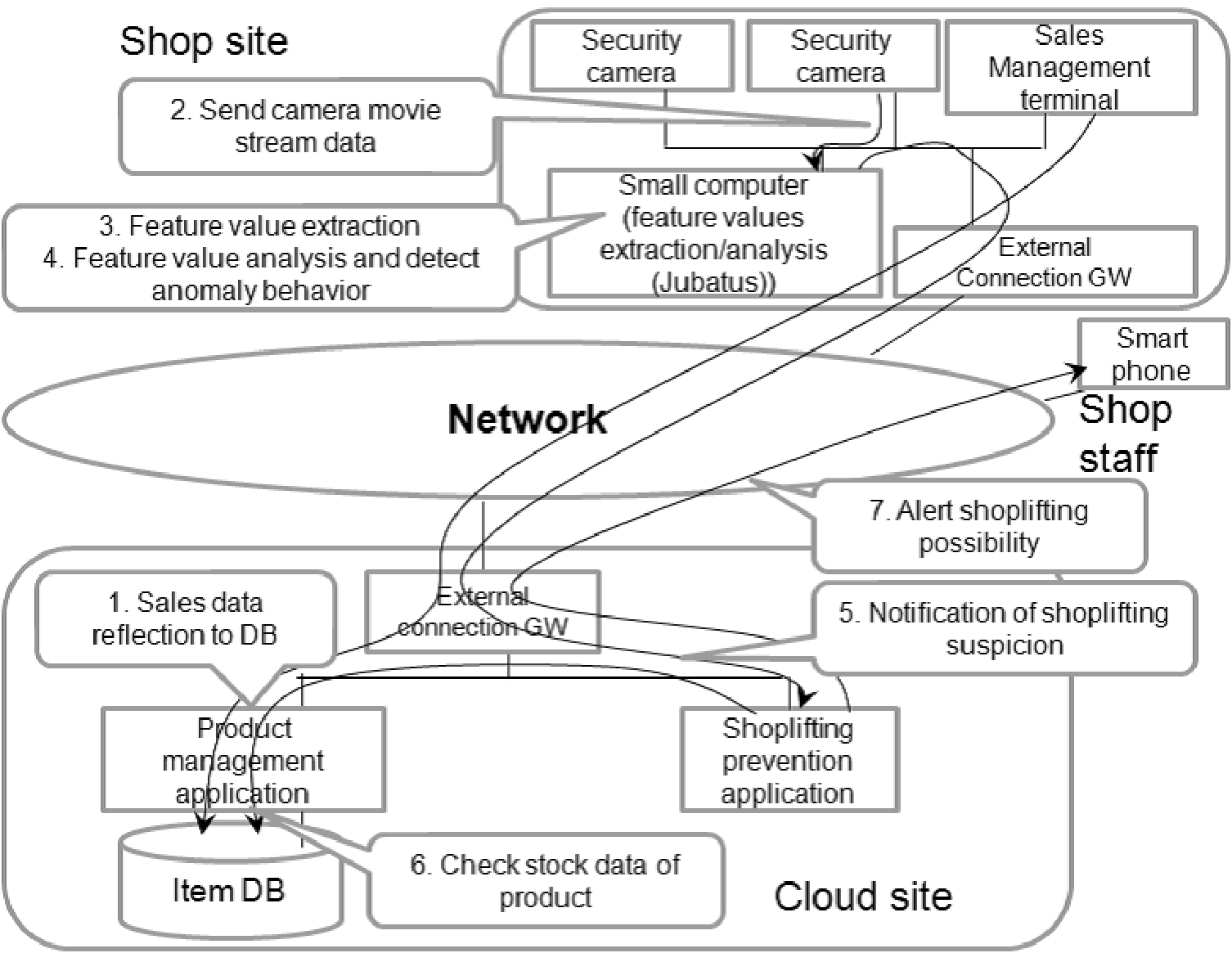}
 \end{center}
 \caption{Proposed system image and processing steps}
 \end{figure}

Step 1: In parallel with shoplifting detection using camera movie, a sales management terminal sends sales information to a product management application on a cloud via a network. A product management application is SaaS which provides business application of ERP, and information of sales and product item stock is stored in item DB. Product item stock information is reflected to item DB in accordance with sales.

Step 2: Stream data of security camera movie is sent to a small computer in a shop. A small computer is a computer which has a certain degree of computation power, memory size and communication capability. For example, Rasbpberry Pi can be used for this to analyze images. 

Step 3: A small computer cuts off each image from movie and extracts feature values from the image data. To extract feature values, libraries of dlib, OpenCV can be used.

Step 4: A small computer detects customer's suspicious behavior from feature values. To analyze stream data of feature values, we use online machine learning Jubatus\cite{Jubatus}. Jubatus can detect not only shoplifting behavior based on pre-defined rules but also suspicious behavior based on machine learning.

Step 5: When Jubatus detects a shoplifting suspicion such as anomaly score is high, image data and related data are sent to a shoplifting prevention application on a cloud. 

Step 6: A shoplifting prevention application checks product item stock data in item DB because image analysis accuracy is not sufficient. If there is a shoplifting, there is inconsistency between stock in item DB and actual stock in a shelf. Actual stock in a shelf also can be detected by security camera image. 

Step 7: A shoplifting prevention application notifies an alert with suspected customer image to smart phones of shop staffs when item DB check leads a high possibility of shoplifting.

%% file: section4E.tex
\section{Confirmation of movie analysis by Jubatus}
We confirmed a precision ratio of security camera movie analysis by Jubatus stream processing. To estimate shoplifting actions, firstly we checked to judge users' posture. We implemented Jubatus plug-in which extracts feature values from an image and Python client which judges users' posture from one image of security camera by Jubatus.

To extract feature values, we used dlib library which extracts 68 coordinate points of eyes, nose, mouth, shape of the face from face images. From 68 coordinate points, we separate to X and Y axis and we obtain 136 feature values. For obtained feature values, we normalize relative coordinate in each face image by deducting face image position and dividing face size. Normalized data is classified by Jubatus classification functions of Linear Classifier and kNN Classifier. We adopt good precision one from two results. (See, Fig.2) 

We trained and judged postures by Jubatus for 1,103 images and verified precision ratio. Precision ratio of k-cross validation was 72\% when k was 10. This test was simple and there remained some non-tuning points. Through this verification test, we confirmed that we could judge customers' posture in a certain degree of precision ratio from security camera movie. It was also said because precision ratio of security camera analysis was not 100\% and some tunings were needed for each shop environment, we needed to enhance accuracy of shoplifting detection by checking item DB of ERP. 

Machine learning models and other configurations of Jubatus are distributed from a cloud shoplifting prevention application to small computers. When we update configurationss, we will use cloud batch updating methods such as \cite{Yamato4} or server coordinating methods such as \cite{Yamato7}\cite{Yamato5}. And when we conduct regression tests after configuration updates, we will use automatic verification methods such as \cite{Yamato2}.

 \begin{figure}[tb]
 \begin{center}
  \includegraphics[width=84mm]{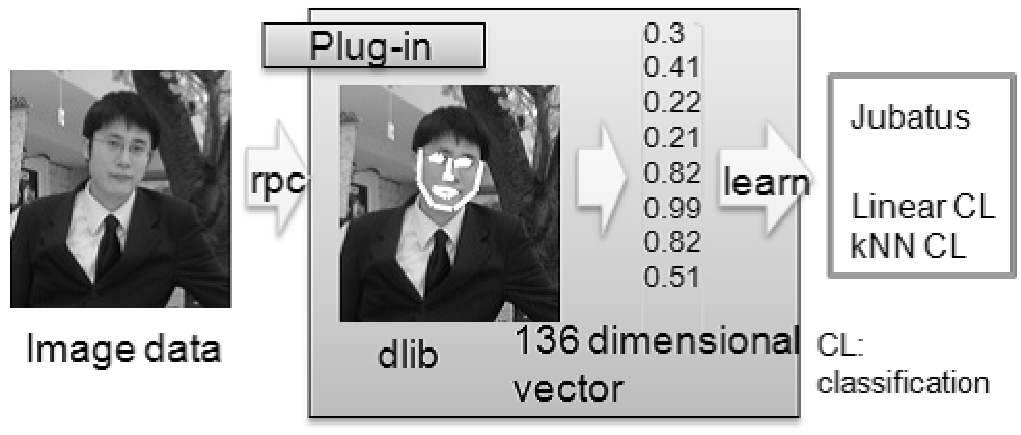}
 \end{center}
 \caption{Test outline of security camera movie analysis by Jubatus}
 \end{figure}

%% file: section5E.tex
\section{Conclusion}
We proposed a low cost shoplifting prevention service for small retail shops. In our proposal, Jubatus on small computers deployed in shop sites analyzed security camera movie, detected anomaly behaviors of customers and notified to a cloud, and a shoplifting prevention application on a cloud checked product item DB in ERP and notified shop staffs by emails when a possibility of shoplifting was high enough.